\documentclass[3p, twocolumn]{elsarticle}

\usepackage{hyperref}
\usepackage{tikz}
\usepackage{mathtools}
\usepackage{color}
\usetikzlibrary{calc}
\tikzset{every picture/.style={font issue=\tiny},
         font issue/.style={execute at begin picture={#1\selectfont}}
        }
\usepackage{pgfplots}
\usepackage{subfig}\usepackage{dialogue}

\usepackage{algorithm,bm}
\usepackage{algorithmic}

\usepackage{filecontents}
\usetikzlibrary{positioning,decorations.pathreplacing,shapes,3d, math}
\usetikzlibrary{calc,patterns,angles,quotes}
\usepackage{colortbl}

\begin{document}

\begin{frontmatter}

\title{High-Energy Density Hohlraum Design Using Forward and Inverse Deep Neural Networks}

\author{Ryan G.\ McClarren}
\address{University of Notre Dame, 365 Fitzpatrick Hall, Notre Dame, Indiana}
\ead{rmcclarr@nd.edu}

\author{I.L.~Tregillis, Todd J.~Urbatsch, E.S.~Dodd}
\address{Los Alamos National Laboratory, Los Alamos, New Mexico}
\ead{iant@lanl.gov, tmonster@lanl.gov}

\begin{abstract}
We present a study of using machine learning to enhance hohlraum design for opacity measurement experiments.  For opacity experiments we desire a hohlraum that, when its interior walls are illuminated by the National Ignition Facility (NIF) lasers, will produce a high radiation flux that heats a central sample to a temperature that is constant over a measurement time window.  Given a baseline hohlraum design and a computational model, we train a deep neural network to predict the time evolution of the radiation temperature as measured by the Dante diagnostic.  This enables us to rapidly explore design space and determine the effect of adjusting design parameters.  We also construct an ``inverse'' machine learning model that predicts the design parameters given a desired time history of radiation temperature.  Calculations using the machine learning model demonstrate that improved performance over the baseline hohlraum would reduce uncertainties in experimental opacity measurements.

\end{abstract}


\end{frontmatter}


\section{Experimental Design for Opacity Measurements}
Hohlraums are used in High Energy Density Physics (HEDP) and Inertial Confinement Fusion (ICF) experiments to convert laser energy to thermal x-rays for imploding capsules, heating targets, and generating thermal radiation waves \cite{drake2018high}.  The physics of lasers, wall heating, and ablation are complex and often not well modeled.  Also, the complete set of physical phenomena that are understood to be present in an experiment are not always included in the simulation codes. Even so, it remains expensive, in terms of computer resources and human development time, to use these codes to model a given experiment.  
Thus, designing hohlraums can be a costly and skillful art that is augmented with simplified modeling and experiments.  Our goal here is to design a machine learning model that can give designers more insight into their hohlraums and lessen the time required to design them.

For the Opacity-on-NIF experimental campaign at the National Ignition Facility (NIF), five hohlraums were designed, simulated, and experimentally shot during the course of a year \cite{dodd2018hohlraum} to produce conditions for measuring the opacity of different material samples.  The goal of that initial design effort was to heat the target at the center of the hohlraum to temperatures of 160-195 eV and electron densities ranging from $0.7 \times 10^{22}$ to $4 \times 10^{22}$ cm$^{-3}$ and to keep the centerline view clear for backlighting and spectroscopy.  The target also had to be in a stable, uniform state during the backlighter measurement and, most importantly, it had to be in local thermodynamic equilibrium (LTE).  As  development of the platform continues toward the higher temperatures, avoiding non-LTE and target gradients becomes increasingly difficult, and it is clear that other considerations, such as emission from the hohlraum contributing to spectrometer background, tamper behavior, and drive dependence, become important.

Despite the Gadarene rush to apply machine learning to all manner of scientific problems \cite{vento2019traps}, the understanding of simulation outputs is a particularly fruitful application. There has been success in the high-energy density physics community in using machine learning to understand the relationships between experiment and simulation. This includes the post-shot analysis of NIF experiments \cite{humbird2019parameter}, transfer learning to obtain better predictions of experimental outputs \cite{humbird2019transfer}, and models to predict ICF fusion yields \cite{Hatfield:2020fk}. 
The goal of our effort is to use data from simulations to train machine learning models to predict the performance of a given hohlraum design in an experiment.  We accomplish this using deep neural networks to learn a {\em forward} model that predicts experimental diagnostics given a hohlraum geometry and laser pulse profile and an {\em inverse} model that given the experimental outputs attempts to infer the hohlraum geometry and laser pulse. In this sense we hope that the models ``learn'' the relationships between design parameters so that the design space can be rapidly searched for candidate designs. These candidate designs can then be fed into the simulation code to confirm the prediction of the machine learning model. 

Focusing on the NOVA-style hohlraum with baffles from the early development of Opacity-on-NIF hohlraums\cite{dodd2018hohlraum}, we describe the set of training simulations for the model in the following section. We then develop a machine learning model to predict the peak temperature of the target and its time profile in Section 3, and discuss the results and characteristics of the models in Section 4. In Section 5 we use the models we develop to design a hohlraum that would improve the uncertainty in a hypothetical opacity measurement.


\section{Simulations of laser-driven hohlraums}

\begin{figure}[h]
\centering
\begin{tikzpicture}[scale=6]
\fill [ball color=yellow, shading=ball] (-0.025,0.2) rectangle (0.025,-0.0);
\draw[->, black!50] (-0.61,0) -- (0.6,0) node(xline)[right,text=black]
        {$Z$};
\draw[-, black!50] (0,.4) -- (0,-.28875*0-0.01) node(yline)[below,text=black]
        [yshift=-1ex]{$Z=0$};
\draw[-, black!50] plot  coordinates {(0.5,0) (0.5,.35) };
\draw[-, black!50] plot  coordinates {(-0.6,0.28875) (0,.28875) };
\draw[-,line width=0.15cm, tension=0.4] plot [smooth] coordinates {(-0.5,.155)  (-.4,.28875)  (0.4,.28875) (0.5,.155)  };
\draw[line width=0.15cm, tension=0.4] plot [smooth] coordinates {(-0.2,.28875+0.01) (-0.2,.1155) };
\draw[-,line width=0.15cm, tension=0.4] plot [smooth] coordinates {(0.2,.28875+0.01) (0.2,.1155) };
\draw[<-, thick] plot  coordinates {(0.2,.1155)  (0.2,0) } node(Ra)[below ,text=black] [align=left]{$\mathrm{R}_{\mathrm{apt}}$=\\0.1155 cm} 
        ;
\draw[<-, thick] plot  coordinates {(0.5,.1550) (0.5,0)  }node(Rleh)[below,text=black]
        [align=left]{$\mathrm{R}_{\mathrm{LEH}}$=\\0.155 cm};
\draw[->, thick]   (-0.6,0)  node(Rleh)[above right,text=black]
        {$\mathrm{R}_{\mathrm{hoh}}$=0.28875 cm}  --  (-0.6,.28875);
\draw[->, thick]   (0,.35)  node(Rleh)[ above right,text=black]
        {$\mathrm{Z}_{\mathrm{hoh}}$=0.5 cm}  --  (0.5,.35);
\draw[->, thick] plot  coordinates {(0.0,0.25) (0.2-0.01,.25) }node(Rleh)[below left,text=black, align=left]
        {$\mathrm{Z}_{\mathrm{baf}}$=\\0.2 cm};
\end{tikzpicture}
\caption{The nominal hohlraum design used in this study. In the terminology of this report this hohlraum has {\tt scale} $=$ {\tt sc\_length} $= R_\mathrm{apt} = 1$. A sample is shown centered at $Z=0$ for illustrative purposes.}
\label{fig:hohlraum_schm}       
\end{figure}
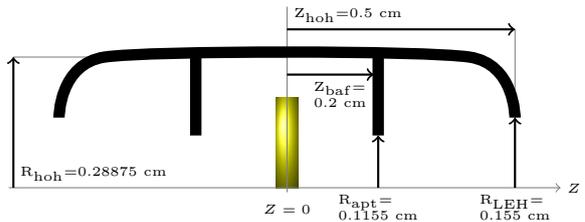
Our simulation strategy followed the procedure detailed in \cite{dodd2018hohlraum}. 
The simulations utilized two laser sources,
representing NIF beam cones oriented at $44.5^\circ$ and
$50^\circ$ from the hohlraum axis. Both source 
descriptions incorporated NIF phase-plate data in order to
reproduce
as closely as possible an empirical beam
intensity pattern in the plane of best focus (i.e.,
transverse to the beam propagation direction). The central elliptical
high-intensity regions had semi-axes $635 \times 367$ $\mu$m and
$593 \times 343$ $\mu$m, while the beam
pointings were such that the points of best focus landed
on the hohlraum axis 180 and 20 $\mu$m outside the
laser entrance hole (LEH) for the $44.5^\circ$ and 
$50^\circ$ beams, respectively. Each laser source delivered 125 kJ to the
target.

\subsection{Synthetic Dante Radiation Temperature Data}

A Dante instrument comprises a multichannel X-ray diode
array for detecting and quantifying time-dependent radiant 
fluxes over a wide range of energies
\cite{Kornblum86, Dewald04, Campbell04, Sorce06, Kline10}.
A critical
diagnostic for many laser hohlraum experiments, Dante is primarily used to obtain 
a measure of the time-dependent radiation temperature
($T_r(t)$). This Dante-derived quantity is frequently
the benchmark against which radiation-hydrodynamic 
hohlraum simulations are 
calibrated \cite{dodd2018hohlraum, Guymer15, Fryer16}. We have
adopted a similar approach for the current study.

The NIF Dante instruments can field up to 18 channels \cite{dodd2018hohlraum, Guymer15}. In aggregate, the channels span a domain covering approximately 20 eV - 20 keV
\cite{Dewald04, Kline10}. The individual channel response
functions typically vary over several orders of magnitude in sensitivity. Narrower channels 
might span 150 eV across the top two decades; wider channels may span several
keV or more. The channel responses overlap significantly.

The raw channel data comprise a set of time-dependent voltage traces.  Given 
the known channel response functions, cable and filter attenuations, and other factors 
\cite{May10, May12}, these traces can be combined to provide a time-dependent 
measure of the bolometric radiant intensity or, subject to blackbody 
assumptions, the radiation temperature within the hohlraum
\cite{Seifter08}. Uncertainties of 
$\pm 7\%$ are not uncommon for measurements of the integrated flux obtained in
this way \cite{Olson12}. 

While it is possible to forward model the individual
channel voltages
using fluxes derived from a radiation-hydrodynamic calculation \cite{Fryer16},
that was unnecessary for the present work. Instead we found it sufficient to 
generate a synthetic time-dependent Dante radiation temperature directly
from the rad-hydro fluxes using a simpler but very common technique.

The postprocessing calculation begins by defining a bundle of parallel rays with an orientation
and field of view approximating that of the Dante instrument 
($37.38^\circ$ from the hohlraum axis for NIF Dante-1). Each ray corresponds to a single pixel
in a simulated detector plane. (This plane is not intended to represent the
Dante instrument itself, but instead functions as a tally surface for computing
post-processed fluxes.) By solving the radiation transfer 
equation along each ray, we obtain the approximate spectral surface brightness 
$I_{\nu}$ seen by each pixel. These brightnesses are  integrated over a 
120 ps window centered on each temporal datapoint, with the resulting time-integrated brightnesses subsequently integrated over energy. By 
computing the bolometric flux over the entire pixel plane (assuming 
orthographic projection along the ray bundle), and equating this to the flux
emitted from a blackbody source, we obtain the blackbody radiation temperature
seen by every pixel in the detector plane.
These operations utilized the Yorick DRAT
library for calculating radiative transfer integrations through a cylindrical
Lagrangian mesh \cite{YORICK}.

The per-pixel radiation temperatures must be averaged or combined in some way 
to obtain a single radiation temperature for each synthetic measurement time.
In this work, we compute two separate aggregate values: an intensity-weighted average, and the area-corrected 
fourth root of the mean intensity where the area correction accounts for the
difference between the rectilinear synthetic detector plane and the 
elliptic projection of the circular hohlraum laser entrance hole. In the
limit of infinitely small pixels and uniform temperature fields, these two averaging methods are known to 
bracket the true radiation temperature \cite{LAUR-14-26986}.
We therefore use the spread between the $T_{r}$ values derived from
these averaging methods as an approximate error bar on the synthetic diagnostic.

Each individual radiation-hydrodynamic simulation, 
corresponding to a different design parameterization as
described below, was post-processed in this fashion to 
obtain a synthetic time-dependent radiation temperature
measurement, $T_r(t)$. These formed the training dataset 
for the current study.

\subsection{Design Parameterization}

\begin{figure}[h]
\centering
\includegraphics[width=0.98\linewidth]{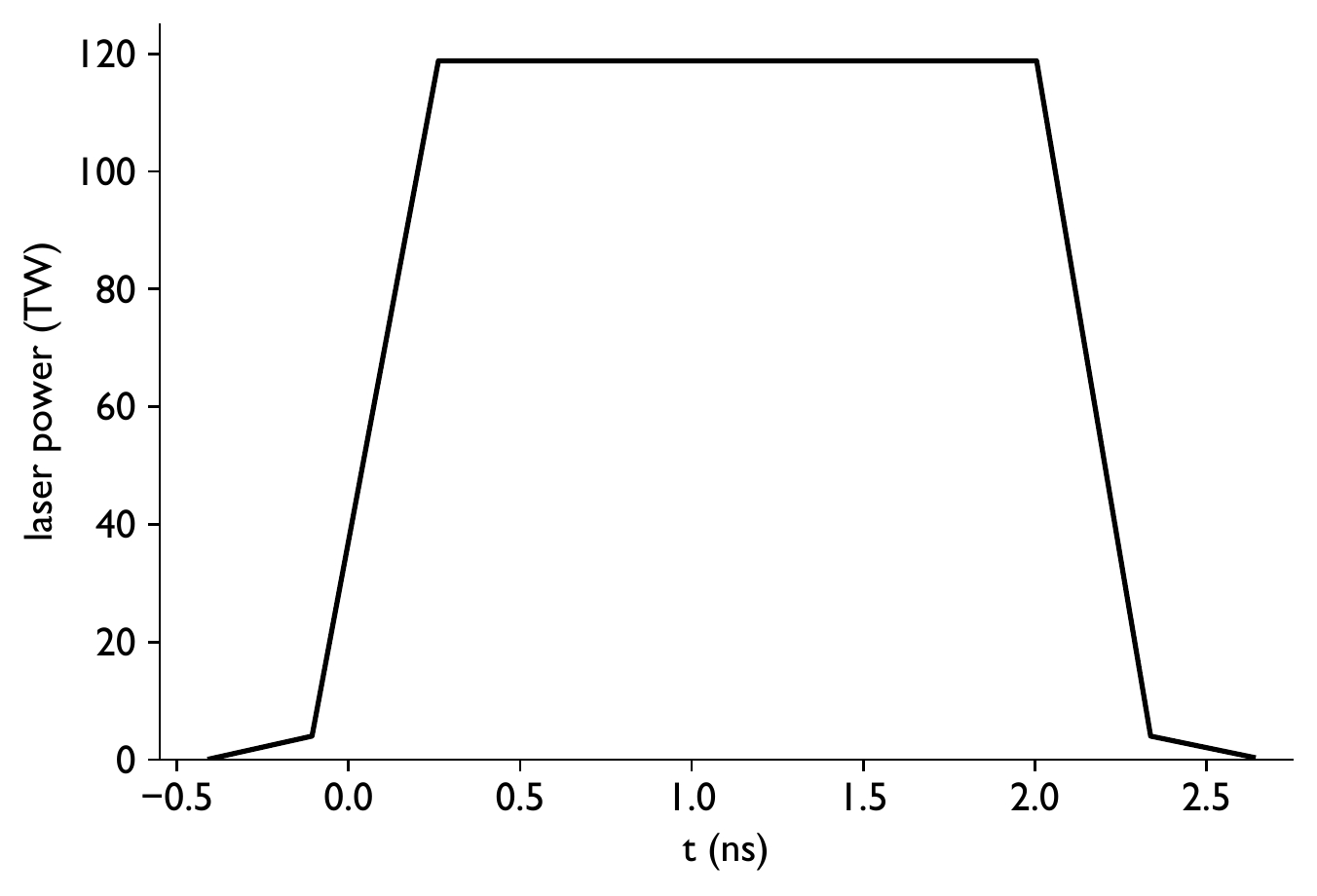} 
\caption{The nominal (\texttt{pulse\_length} = 1) laser drive in the rad-hydro simulations delivered 250 kJ to the target, split evenly between the $44.5^\circ$ and $50^\circ$ beams, over 3 ns.}
\label{fig:pulse}       
\end{figure}

Four multiplicative factors defined the suite of variations applied to
the nominal hohlraum (see Fig.~\ref{fig:hohlraum_schm})
in this study. These were:

\begin{itemize}
\item An overall {\tt scale} parameter for setting the hohlraum size. Except for wall thickness, which did not change in these variations, every dimension in the geometry ($R_\mathrm{hoh}$, $R_\mathrm{apt}$, $R_\mathrm{LEH}$, $Z_\mathrm{hoh}$, and $Z_\mathrm{baf}$) was proportional to this factor.  
\item  An ({\tt sc\_length}) (``sample chamber length'') parameter used to vary $Z_\mathrm{baf}$ while keeping the ratio $R_\mathrm{apt} /Z_\mathrm{baf}$ a constant. 
\item An $R_\mathrm{apt}$ (``aperture radius'') parameter for perturbating the size of the aperture between the sample and laser illumination chambers \emph{independently} of $Z_\mathrm{baf}$. 
\item A {\tt pulse\_length} scaling parameter, used for changing the temporal duration of the laser pulse drive 
(see Fig.~\ref{fig:pulse}) without altering the delivered energy of 250 kJ.
\end{itemize}

Each parameter was varied over the range 0.8 to 1.25 at the levels 0.8, 0.85, 1.0, 1.05, 1.2, and 1.25. When referring to the parameters for a simulation we write the parameters as the vector ({\tt scale},{\tt sc\_length},$R_\mathrm{apt}$, {\tt pulse\_length}). 
The nominal hohlraum corresponds to the set of values 
$\left\{1.0, 1.0, 1.0, 1.0\right\}$.

\begin{figure}
    \includegraphics[width=0.88\columnwidth]{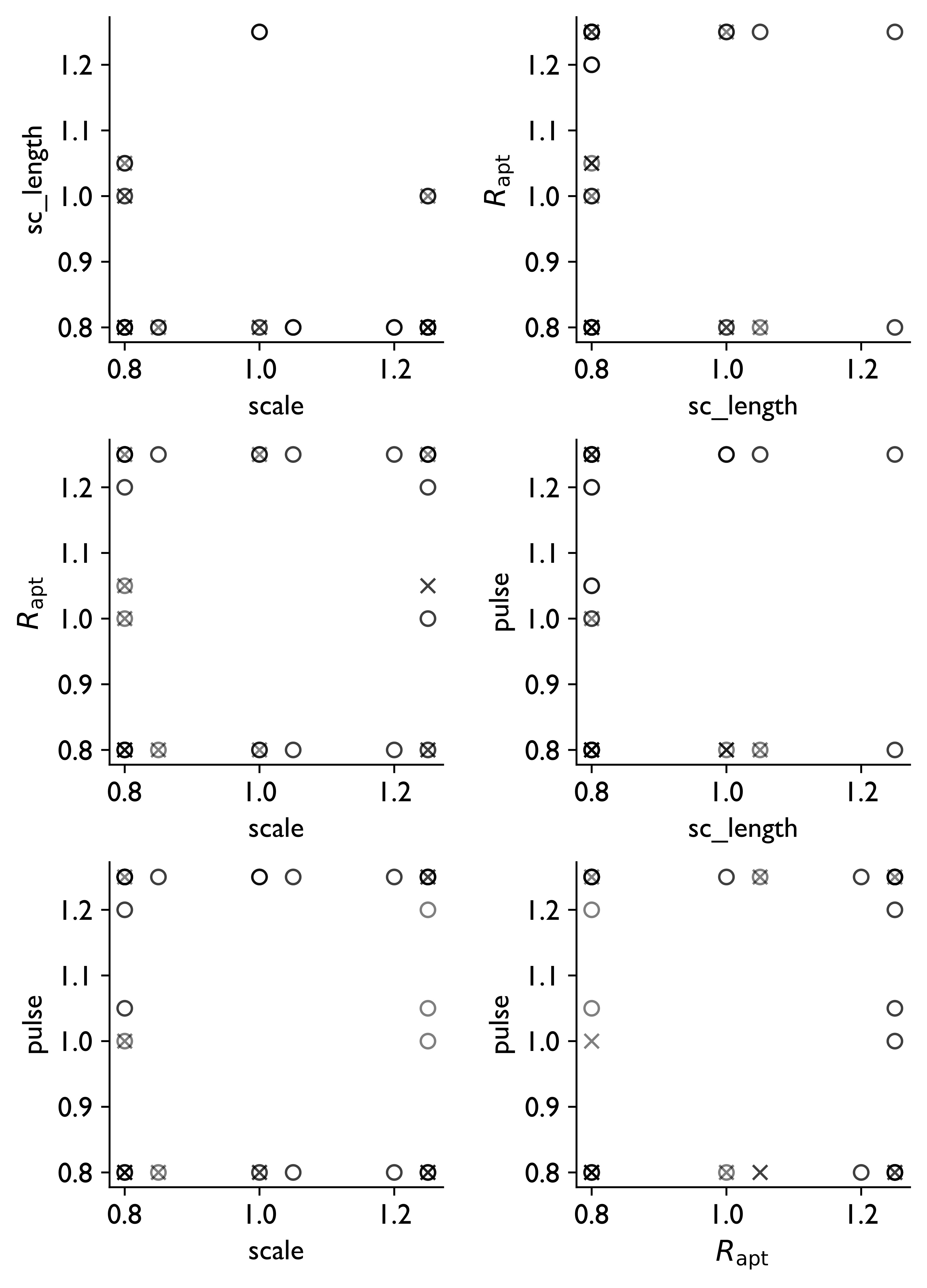}
    \caption{Values of the 4 design parameters in the simulation ensemble. Circles indicate simulations that completed, X's denote runs that failed.}
    \label{fig:runset}
\end{figure}

The parameter combinations utilized for this study are shown in Figure~\ref{fig:runset}. The simulations selected for our ensemble simulations, or run set, were at the corners of the 4-D hypercube of the parameter space for all of the parameters except the sample chamber length, and at the middle of each face of this cube. In total, there were $46$ simulations that ran to completion. The simulations that failed are denoted by an ``X'' marker in Figure~\ref{fig:runset}.

\section{Forward and Inverse Deep Neural Networks}
From each simulation in the ensemble the output we receive is the Dante measurement of the radiation temperature as a function of time, $T_\mathrm{D}(t;\theta)$, where $\theta = (\mathtt{scale},\mathtt{sc\_length}, R_\mathrm{apt}, \mathtt{pulse\_length})$. The times output by the simulation are not at equally-spaced intervals; therefore, we use cubic spline interpolation to put them at a unified set of 50 time points. For training the model we also scale the time variable by the inverse of the laser pulse scale factor so that the time variable gives a  position relative to the laser pulse.

We use a feed-forward neural network to approximate the function $T_\mathrm{D}(t;\theta)$ at the 50 time points. The network we use has six layers: a dense layer with 8 hidden units, a dense layer with 100 hidden units using dropout with a drop probability of 0.05, a dense layer with 50 hidden units, and a convolution layer with a kernel of size 3.  We trained this network using Tensorflow \cite{tensorflow2015-whitepaper}.  Due to the small size of the training data, we performed a leave-one-out cross-validation \cite{mcclarren2018uncertainty} to estimate the mean-absolute error in the model to be about 0.003 keV.

We also desire to have an inverse model that can infer hohlraum design and laser pulse length from the radiation temperature as measured by the Dante diagnostic.  This would allow an experiment designer to specify the desired radiation temperature as a function of time and then determine what hohlraum/laser pulse will give that combination. These inverse problems are challenging because they are likely to be ill-posed in that there are multiple parameters that can give the same temperature time history. It has recently been shown that combining forward and inverse models, one can accurately obtain simulation parameters for an inertial confinement fusion experiment using only diagnostic outputs \cite{humbird2017using}. The models we develop here can be utilized in a similar manner.

The inverse model will take the radiation temperature at the 50 time points as input and then produce the four simulation parameters as output.  We use a neural network of three convolution layers with 2, 4, and 8 kernels of size 2 in the layers, with the output of the last of these layers being a vector of length 392.  Next, ``dropout'' is applied to this vector using softplus activation functions and a dropout probability of 0.05.  Typically, dropout provides robustness to overfitting by setting to zero, or dropping out, certain connections in the network during training.  After dropout, we pass the results through a layer of 4 neurons to produce the final output. 
We also use dropout during the prediction phase to get a distribution on our simulation parameters \cite{humbird2018deep}.



\section{Results}
\subsection{Forward model}
\begin{figure}
    \subfloat[varying scale]{\label{fig:scale}\includegraphics[width=0.88\columnwidth]{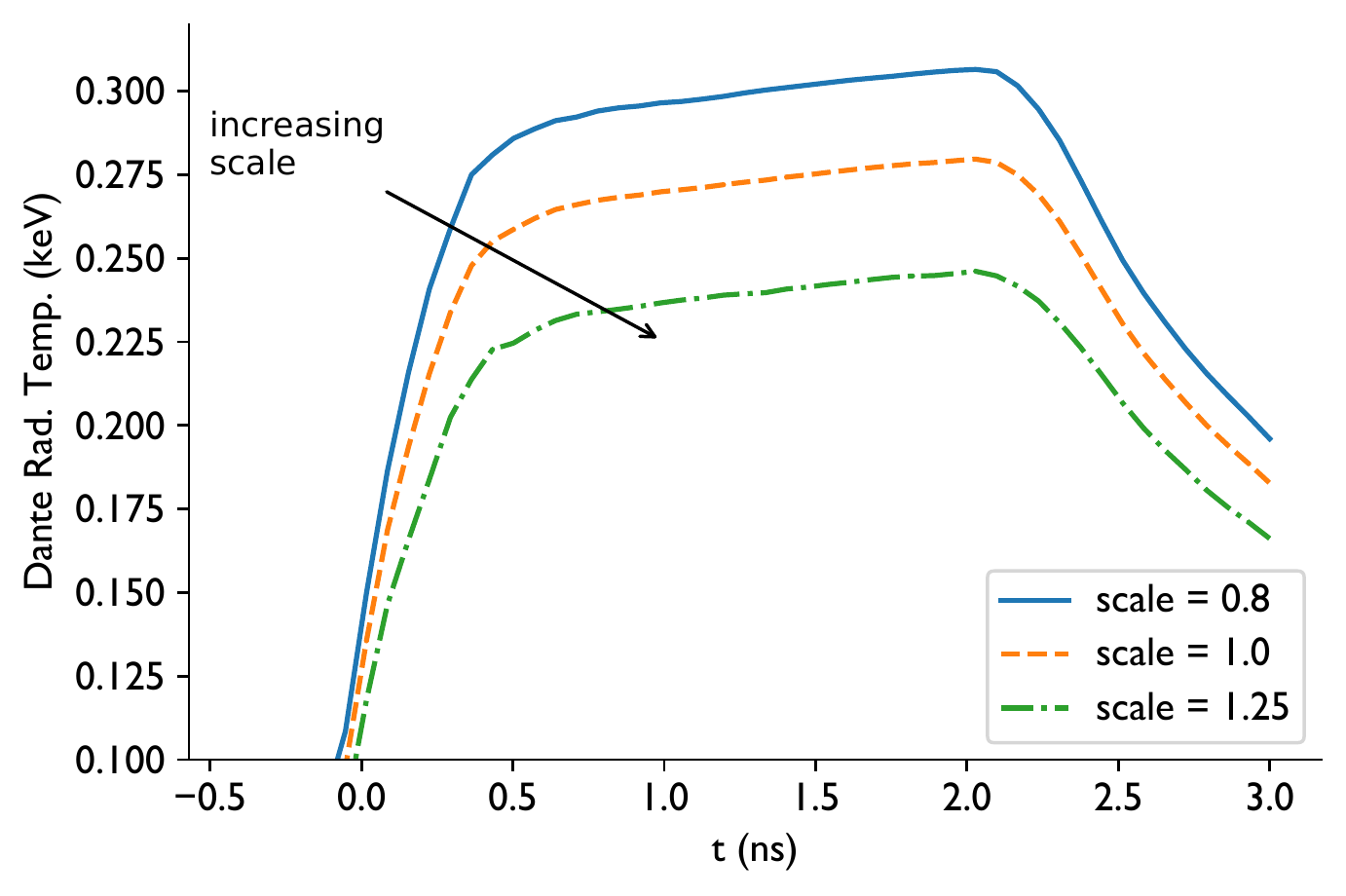}}\\
    \subfloat[varying sample chamber length ]{\label{fig:sample chamber}\includegraphics[width=0.88\columnwidth]{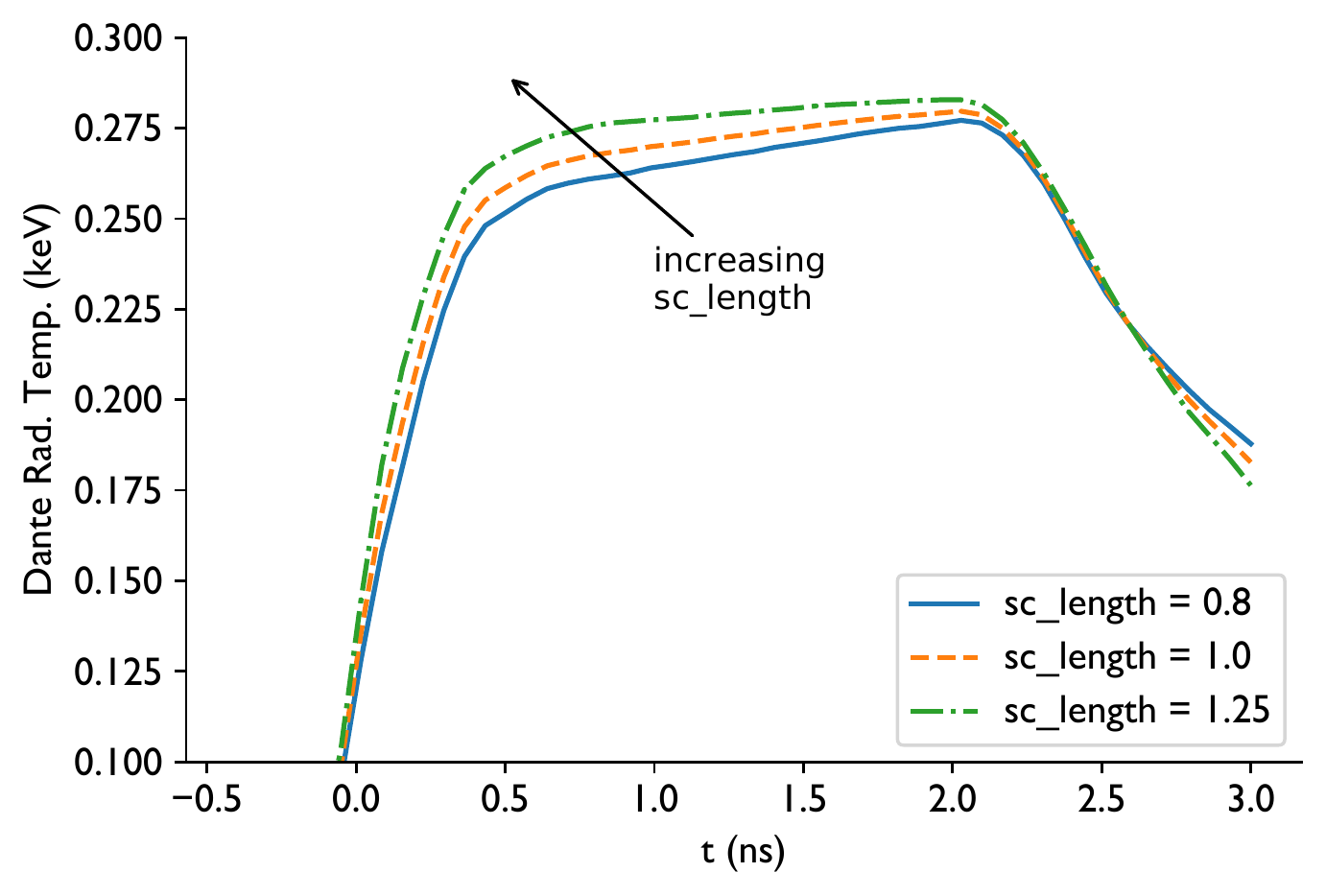}}\\
    \subfloat[varying laser pulse length ]{\label{fig:laser}\includegraphics[width=0.88\columnwidth]{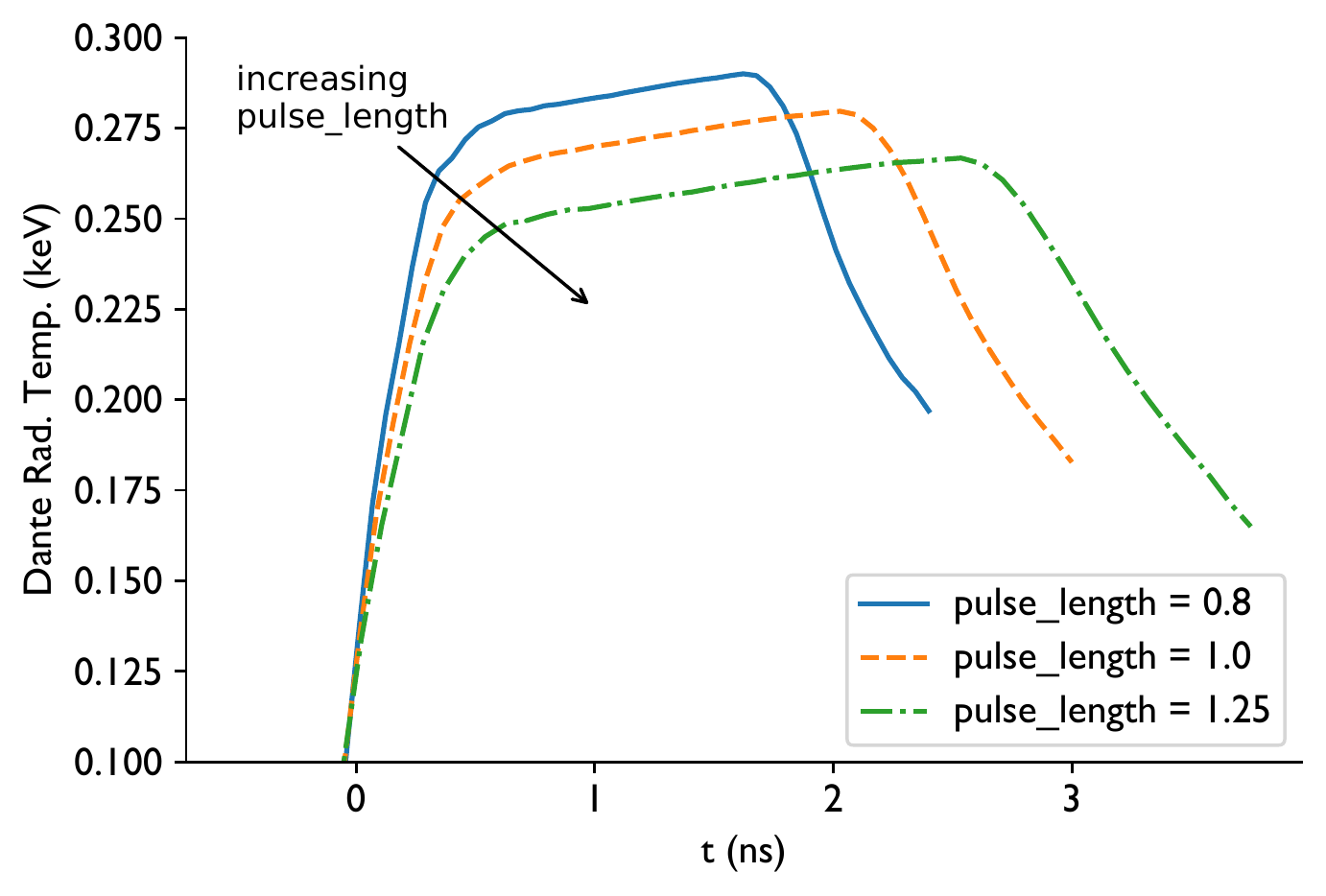}}\\
    \caption{Demonstration of the effect of varying the hohlraum shape parameters in the forward neural network model with every other parameter set to the nominal value of 1.0.  }
    \label{fig:vary}
\end{figure}
From the forward model we can investigate how changing the hohlraum shape parameters and laser pulse affects the radiation temperature as a function of time. In Figure \ref{fig:vary} we show  the  effects of scale, sample chamber length, and laser pulse length. The effect of aperture radius, $R_\mathrm{apt}$, is not shown in the figure because the radiation temperature sampled by the Dante-1 field of view  is effectively independent of this parameter. This is to be expected, as the aperture controls radiation exposure to the sample chamber, whereas the Dante-1 pointing samples the illumination chamber. 
From these figures we observe that decreasing the {\tt scale} parameter has the largest effect on the Dante response by increasing the maximum temperature.
 This is in keeping with our expectations.  Increasing  the {\tt sc\_length} parameter  gives the Dante response a flatter and higher peak temperature. The faster rise to the plateau temperature with increasing {\tt sc\_length} represents the fact that larger {\tt sc\_length} values effectively reduce the volume of the illumination chamber (so temperature rises more quickly). However, because the ratio $Z_\mathrm{baf}/R_\mathrm{apt}$ is kept constant, the aperture radius increases so the illumination chamber temperature doesn't level off to a significantly higher temperature. Both laser pulse length and scale produce an inverse effect on the temperature, and a large laser pulse length also shortens the duration 

\begin{figure}
    \includegraphics[width=0.88\columnwidth]{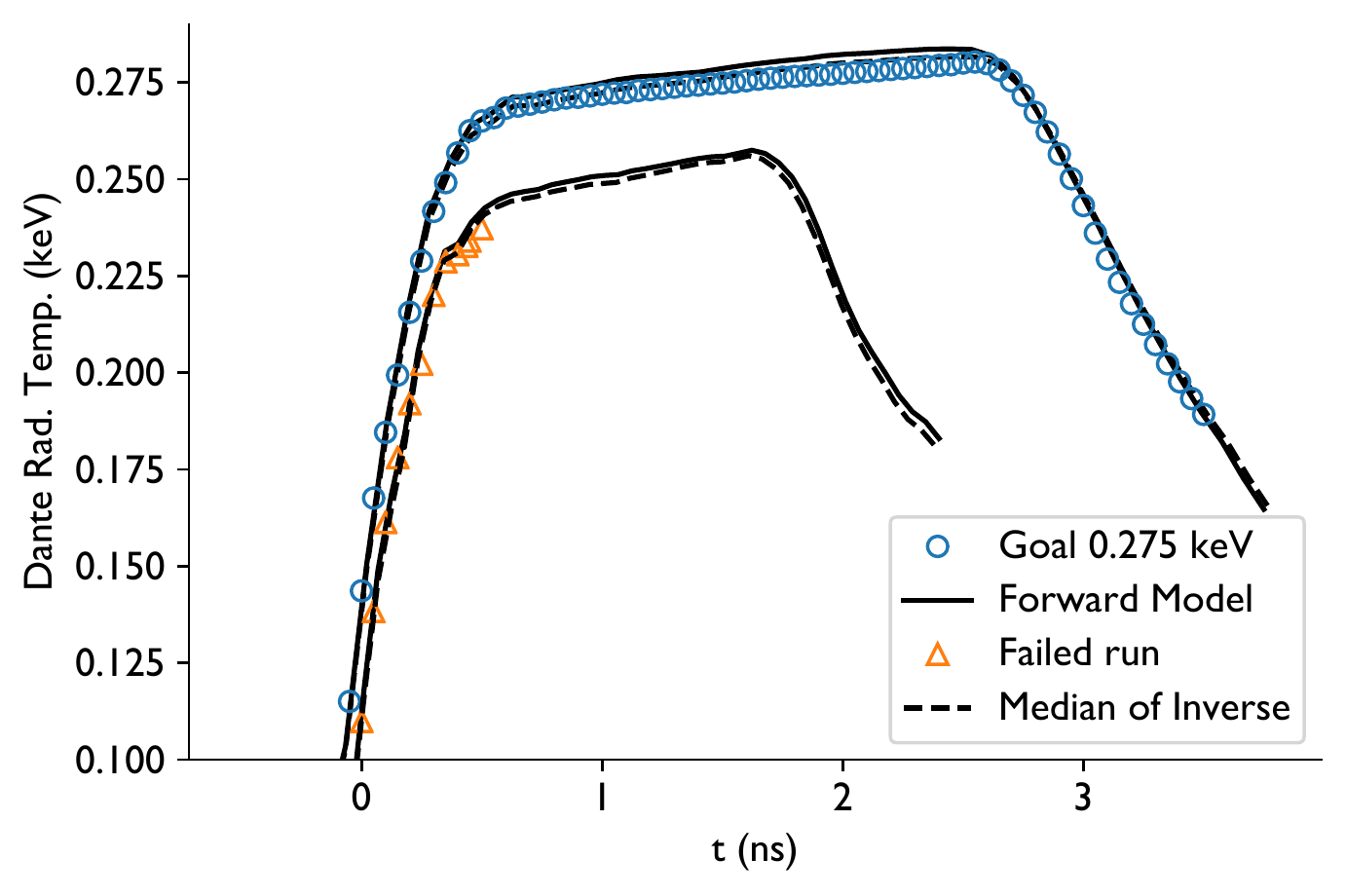}
    \caption{Validation of the forward model on simulations 1) that failed and  2) parameters selected to produce a flat temperature profile of 0.275 keV. }\label{fig:valid}
\end{figure}

In Figure \ref{fig:valid} we validate our results by comparing simulation results to the model prediction for cases not in the training data. The first is a simulation that failed due to mesh tangling at less than 1 ns with parameters $(1.25,1,0.8,0.8)$.  For this simulation, the neural network model gives a prediction that matches the simulation at early times and can estimate the complete history for that simulation.  We also compare to a simulation chosen to make a flat profile at $0.275$ keV. This simulation had parameters $(0.89642857,1.25,1.05714286,1.25)$. We observe near perfect agreement between the simulation and the model's prediction, which demonstrates that the model will be able to predict hohlraum performance for designs not included in its training data.

\subsection{Inverse model}
\begin{figure}
    \centering
    \includegraphics[width=\columnwidth]{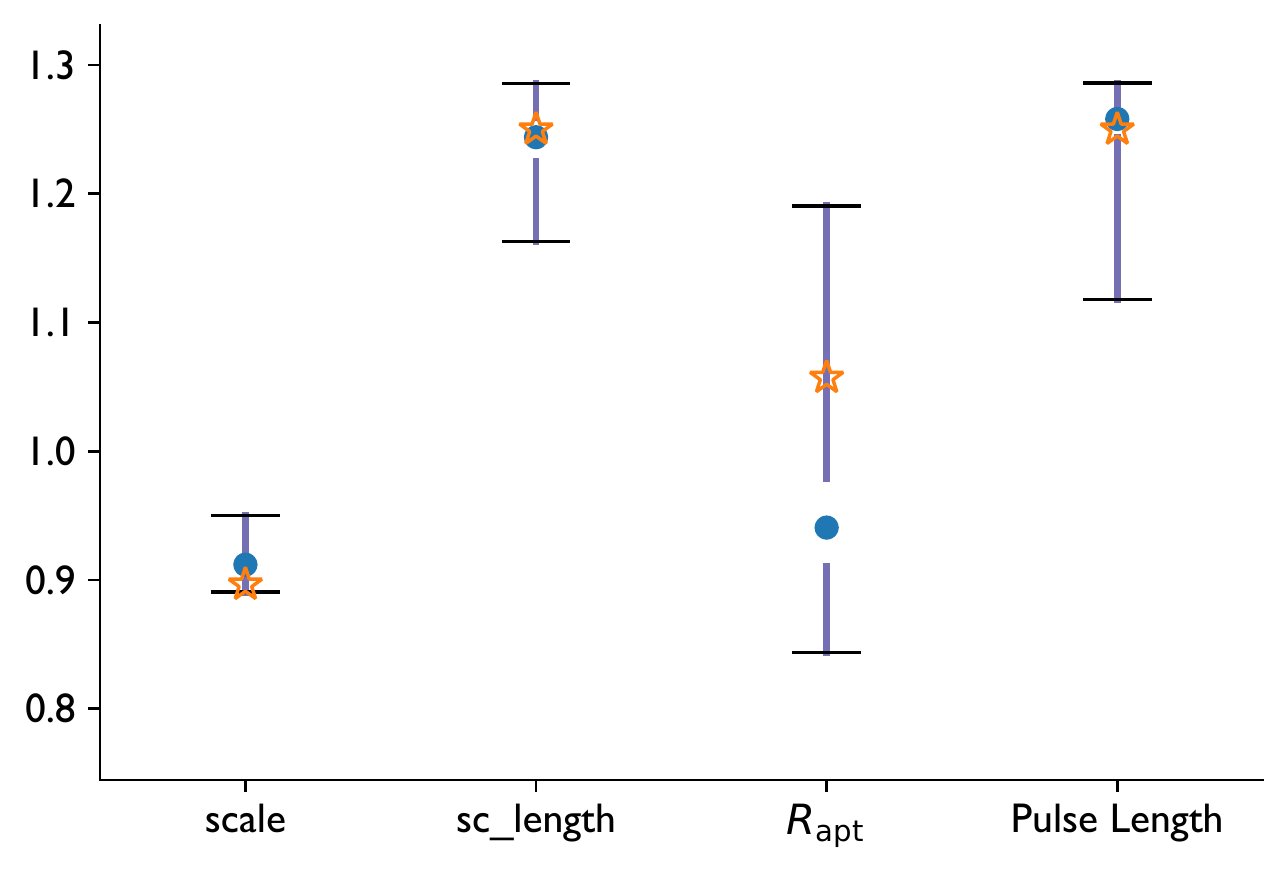}
    \caption{Results from 1000 evaluations of the inverse model when inferring the parameters from the Dante values from a validation simulation with parameters (0.89642857,1.25,1.05714286,1.25). The star symbols indicate the actual values, the dots are the median of the inverse model, and the error bars denote the 95\% confidence interval around the median.}
    \label{fig:inversebox}
\end{figure}

The inverse model allows one to specify a target Dante temperature profile and find the design parameters that will produce the desired response. To test this idea we take the response from the simulation with parameters $(0.89642857,1.25,1.05714286,1.25)$ and input it to the inverse model.  We evaluate the inverse model 1000 times to get a distribution of the input parameters that are consistent with that output. The median and 95\% confidence intervals from these 1000 evaluations, and the actual value that produced the Dante temperature response, are shown in Figure \ref{fig:inversebox}. From this figure we see that for the important parameters in the simulation the median value from the inverse model is close to the true value from the simulation and that the 95\% confidence intervals contain the correct values. We can also understand the range of the confidence intervals as an indicator of how important a parameter is: the {\tt scale} parameter has the smallest confidence interval and  the unimportant $R_\mathrm{apt}$ parameter has a large confidence interval, indicating that it does not matter how we set this parameter.
\section{Application of the Model to Opacity Measurements}
The motivation for this  hohlraum model was to guide the geometric design of the McFee-Apollo hohlraum to achieve increasingly extreme temperatures for the Opacity-on-NIF campaign \cite{perry2017replicating,perry2020progress}, though the present work deliberately uses a simpler design as a proof of concept.  There are more considerations in obtaining quality opacity data, the main being the reduction of backgrounds to the spectrometer (e.g, emission from the hohlraum), improving the signal, and reducing uncertainties in the opacity measurement.  One interesting feature that emerged from this study is the time-dependence of the hohlraum temperature during the backlighter pulse that would measure the opacity.  Note in Figure \ref{fig:sample chamber} how the slope of the plateau is shallower for larger sample chamber lengths and correspondingly shorter baffles.  The main effect is that the interior can heat up faster and then remain closer to its max temperature with a bigger, more open sample chamber.  A competing factor that we do not directly consider here is that these higher interior conditions may be less Planckian due to more direct exposure to the laser hot spots.   In opacity measurements, the sample condition must be nearly constant during the backlighter pulse of 1 ns or less \cite{perry2020progress}.  If we simplistically equate the sample temperature to the hohlraum temperature in Figure \ref{fig:sample chamber}, the flatter plateau is more desirable.  Let us assume further that the sample is a constant density of 0.04 g/cm$^3$ \cite{perry2017replicating}.  Then, what are the ranges of opacities for our material of interest, iron?  Let us consider the full range of the plateau, about 0.5 to 2.0 ns, with the {\tt sc\_length} = 0.8 having a temperature range of about 250-270 eV, and the {\tt sc\_length}=1.25 a range of 270-280 eV.  Presumably, the smaller range is more desirable.

\begin{figure}
    \centering
    \includegraphics[width=\columnwidth]{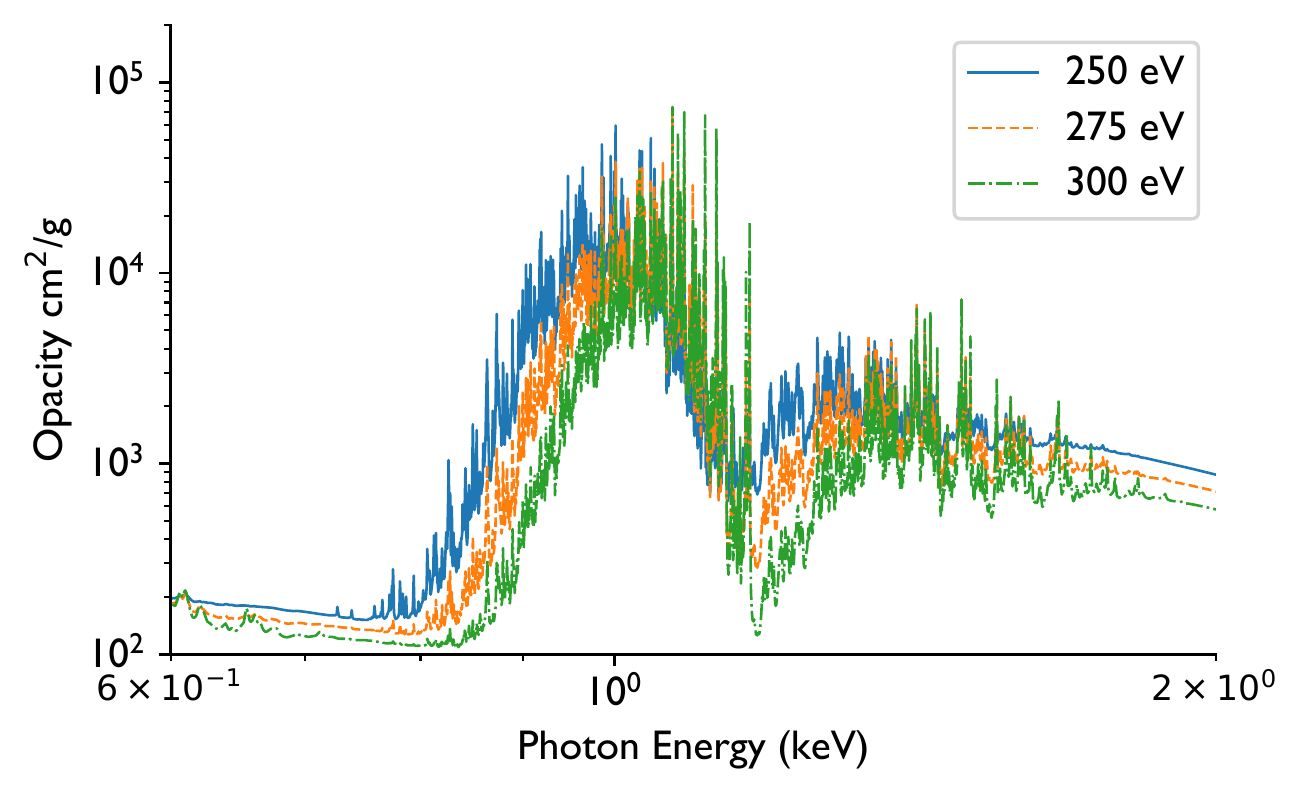}
    \caption{Iron opacity at three different temperatures and a density of 0.04 g/cm$^3$. }
    \label{fig:IronOpacity}
\end{figure}

Then, what are the ranges of opacities for iron, our material of interest?  Utilizing the atomic physics interface at \url{https://aphysics2.lanl.gov} with iron table n12143, we plot in Figure \ref{fig:IronOpacity} the opacities for the nearest available bracketing temperatures of 250, 275, and 300 eV for the energy range of 500-2000eV of interest to the Opacity-on-NIF campaign.  We are more interested in the windows, or valleys, in the opacity, because those are what the radiation will see most. That windowing is dominant in the generation of gray and multigroup opacities using the Rosseland average, which weights the valleys more.  Thus, let us characterize the minimum opacities immediately below and immediately above 1keV.  If we linearly interpolate to 270 and 280eV using nearest neighbors, we have the opacity minima in Table \ref{tab:minopac}.
\begin{table}[]\arrayrulecolor{black!20}
\caption{Location and value of the two local minimum opacities at a range of material temperatures. All temperatures and energies are in eV; opacities are in cm$^2$/g.}\label{tab:minopac}
\begin{center}
\begin{tabular}{c|cccc}
$T$ & $E_1$ & $\sigma_1$  & $E_2$ & $\sigma_2$\\
\hline
250       & 853        & 48.7      & 1183       & 39.8 \\
270       & 882        & 40.7      & 1185       & 24.9 \\
275       & 889        & 38.7      & 1186       & 21.2 \\
280       & 906        & 38.3      & 1207       & 21.0 \\
300       & 972        & 36.5      & 1290       & 20.0
\end{tabular}
\end{center}
\end{table}

For the smaller temperature range, $\Delta T_R$=10eV versus 20 eV, the $\Delta \sigma$ is indeed about one-third as much, as desired and expected.  However, the energy locations of those minima in 270-280eV change as much or more than in 250-270eV, which would tend to increase the uncertainty in the opacity measurement.  Notice in Figure \ref{fig:IronOpacity} the non-monotonic temperature dependence in the 500-700eV spectral range.  

Thus, future machine-learning models may incorporate stability of the opacity sample, and they may incorporate theoretical opacity data, appropriately reduced as necessary, to optimize uncertainty in the experimental opacity measurement.


\section{Conclusion}
We have demonstrated the training and use of forward and inverse deep neural network models to predict hohlraum behavior and to design hohlraums for HEDP and ICF experiments.  We also analyzed iron opacity data for the modeled conditions and motivated a future inclusion of opacity data into the models.

\section*{Acknowledgement}
This work was supported by the US Department of Energy through the Los Alamos National Laboratory. LANL is operated by Triad National Security, LLC, for the National Nuclear Security Administration of the US DOE (Contract No. 89233218CNA000001), LA-UR-20-26126.

\bibliography{mybibfile}

\end{document}